\documentclass[prl,twocolumn]{revtex4}
\usepackage{amsmath,amssymb,amsthm} 
\usepackage{times}
\newcommand{\beq}{\begin{equation}}
\newcommand{\eeq}{\end{equation}}
  \newcommand{\beql}[1]{\begin{equation}\label{eq:#1}}
  \newcommand{\beqa}{\begin{eqnarray}}
  \newcommand{\eeqa}{\end{eqnarray}}
\newcommand{\beqas}{\begin{eqnarray*}}
\newcommand{\eeqas}{\end{eqnarray*}}
\newcommand*{\de}{\delta}
 \newcommand*{\ps}{\psi}                                     
  \newcommand*{\Eq}[1]{Eq.~(\ref{eq:#1})}                                     
  \newcommand*{\eq}[1]{(\ref{eq:#1})}
\newcommand*{\ket}[1]{|#1\rangle}
\newcommand*{\bracket}[1]{\langle#1\rangle}
\newcommand{\ep}{\varepsilon}
\newcommand{\et}{\eta}

\newcommand{\kxi}{\ket{\xi}}

\begin{document}
\title{Comment on ``Proof of Heisenberg's error-disturbance principle''}
\author{Masanao Ozawa}
\affiliation{Graduate School of Information Science,
Nagoya University, Chikusa-ku, Nagoya, 464-8601, Japan}
\begin{abstract}
Recently, Kosugi [arXiv:1504.03779v2 [quant-ph]] argued that 
Heisenberg's error-disturbance relation (EDR) must be interpreted 
as being between the resolution,
the preparational error for the post-measurement observable, and the disturbance.
He further claimed that Heisenberg's EDR can be proven to hold true in general, 
when the meter observable is modified as one of its functions.
Here, some comments are given to suggest that the above claims are not supported.
\end{abstract}
\maketitle

Consider a general measuring process
in which the position $x$ of a mass with the momentum $p$ 
is measured by a measuring interaction 
with a probe having the meter observable $X$
from time 0 to time $t$
such that the measurement outcome is obtained by measuring $X$ at time $t$.
In the Heisenberg picture 
the measurement error (or precision) $\ep(x)$,
the preparatinal error (or resolution, or predictive error) $\de(x)$,
and the momentum disturbance $\et(p)$ of this measurement are defined by 
\beqas
\ep(x)&=&\bracket{\ps,\xi|[X(t)-x(0)]^2|\ps,\xi},\\
\de(x)&=&\bracket{\ps,\xi|[X(t)-x(t)]^2|\ps,\xi},\\
\et(p)&=&\bracket{\ps,\xi|[p(t)-p(0)]^2|\ps,\xi},
\eeqas
where $\ket{\psi}$, $\kxi$,  and $\ket{\psi,\xi}$ stand for the initial states of the object, 
the probe, and the object-probe composite system, respectively 
\cite{88MS,89RS,90QP,91QU,App98a}.

In Ref.~\cite{Kos15}, Kosugi recently argued that Heisenberg's original 
error-disturbance relation (EDR)
\cite{Hei27}
\beql{Hei27}
\ep\et\sim h
\eeq
derived by the $\gamma$-ray microscope thought experiment
should be reformulated as a relation
between the preparational error (the ``resolution'') $\de(x)$, 
instead of the measurement error (the ``precision'') $\ep(x)$, 
and the disturbance $\et(p)$.
He also claimed that the relation 
\beql{RDR}
\de(x)\et(p)\ge\frac{\hbar}{2}
\eeq
always holds if the meter observable $X$ is modified as one of its functions $f(X)$.
From the above he concluded that Heisenberg's original EDR holds true in general,
in contrast to the recent researches 
\cite{03UVR,03HUR,03UPQ,Hal04,WHPWP13,Bra13}
on universally valid error-disturbance relations 
and their experimental confirmations
 \cite{12EDU,RDMHSS12,13EVR,RBBFBW14}.
 
It should be noted that Appleby \cite{App98a} previously proved \Eq{RDR}
under the unbiasedness condition
\beql{UBC}
\bracket{\xi|X(t)-x(t)|\xi}=0
\eeq
as the predictive error-disturbance relation for predictively unbiased measurements,
where $\bracket{\xi|\cdots|\xi}$ stands for the partial inner product.
Kosugi claimed without proof that \Eq{UBC} can be satisfied by replacing $X(t)$ 
by a function $f(X(t))$ of $X(t)$ \footnote{Eq~(12) in Ref.~\cite{Kos15} 
used in the above argument was previously obtained in page 13 of Ref.~\cite{91QU}.}.
However, this is not possible in general.

The claim that Heisenberg's original EDR must be interpreted as relation \eq{RDR}
for the ``resolution'' conflicts with a common view that \Eq{Hei27} leads to the impossibility of 
precise position measurements without disturbing momentum.
In fact, there is a position measurement that satisfies the Born formula in any states 
but has bad ``resolution''.
In contrast note that $\ep(x)$ is determined by the POVM of the measurement, 
that $\ep(x)=0$ if the measurement is precise in the sense that $X(t)$ and $x(0)$ are
perfectly correlated in the state $\ket{\ps,\xi}$ \cite{05PCN,06QPC},
and that $\ep(x)=0$ for all $\ket{\psi}$ if and only if  the measurement outcome
satisfies the Born formula in any states $\ket{\psi}$ \cite{03UVR,03HUR}.

\end{document}